\documentclass[a4paper,11pt]{article}
\pdfoutput=1 

\usepackage{jinstpub} 

\usepackage{amsmath}
\usepackage{amsfonts}
\usepackage{amssymb}

\usepackage{lineno}

\title{Measurement of Liquid Argon Scintillation Light Properties by means 
of an Alpha Source placed inside the CERN 10-PMT LAr Detection System}

\collaboration[c]{for the ICARUS/NP01 Collaboration}

\author[a]{B.~Ali-Mohammadzadeh,}
\author[b,c]{M.~Babicz,}
\author[a]{V.~Bellini,}
\author[d]{A.~Fava,}
\author[c]{U.~Kose,}
\author[c,e]{F.~Pietropaolo,}
\author[f]{M.C.~Prata,}
\author[f,1]{G.L.~Raselli\note{Corresponding author.},}
\author[c]{F.~Resnati,}
\author[f]{M.~Rossella,}
\author[f]{C.~Scagliotti,}
\author[a]{F.~Tortorici}
\author[c,2]{and A.~Zani\note{now at University of Milano and INFN, Milan, Italy}}

\affiliation[a]{University of Catania and INFN, Catania, Italy}
\affiliation[b]{Institute of Nuclear Physics PAN, Cracow, Poland}
\affiliation[c]{CERN, Geneva, Switzerland}
\affiliation[d]{Fermi National Laboratory, Batavia IL, USA}
\affiliation[e]{University of Padova and INFN, Padova, Italy}
\affiliation[f]{University of Pavia and INFN, Pavia, Italy}

\emailAdd{gianluca.raselli@pv.infn.it}

\abstract{
A particle detection system that exploits the scintillation light produced by ionizing particles in liquid argon (LAr) has been assembled at CERN. The system is based on 10 large-area photomultiplier tubes (PMT) immersed in a 1500-liter dewar
filled with liquid argon and equipped with an extendible feed-through and mechanical
support for an alpha source ($^{241}$Am). The position of
the source can be changed with respect to the PMT plane in vertical and horizontal directions. 
Arrays of silicon photomultiplier (SiPM) photodetectors, integrated in the source support, are used for the data acquisition trigger and to define the $t_0$ of the light generation. PMT and SiPM signals can be recorded at different distances and different positions allowing the measurement of some of the LAr scintillation light properties. The system was studied and characterized in detail, and physics
results on LAr scintillation properties are expected soon. 
}


\proceeding{15$^{\text{th}}$ Topical Seminar on Innovative Particle and Radiation Detectors (IPRD19) \\
  14-17 October 2019\\
  Siena, Italy}

\begin{document}

\maketitle
\flushbottom

\section{Introduction}

The detection of light produced by the scintillation of the liquid argon (LAr) plays a crucial role for triggering and for the determination of the absolute time of events in experiments exploiting the Liquid Argon Time Projection Chamber (LAr-TPC) technique for neutrino physics and rare events search.

A LAr test facility, set up at CERN, is used to perform small-scale studies on performances of data acquisition (DAQ) and trigger electronics to be installed in larger volume LAr-TPCs, such as ICARUS T600 at FNAL (see~\cite{Raselli:2019tnp} and references therein).
Ionizing events were already measured by exposing the system to cosmic rays~\cite{BABICZ2019162421}. The facility is now instrumented with an alpha source mounted on an extendible mechanical handler which permits to vary the source position inside the active volume.

The main characteristics of the test facility equipped with the alpha source are described in Section 2
of this paper, while the triggering and data taking are presented in Section 3. 
A preliminary data analysis and the characterization of the system
are reported and discussed in Section 4.

\section{The CERN 10-PMT facility}

\begin{figure}[!t]
\centering
\includegraphics[width=0.5\columnwidth]{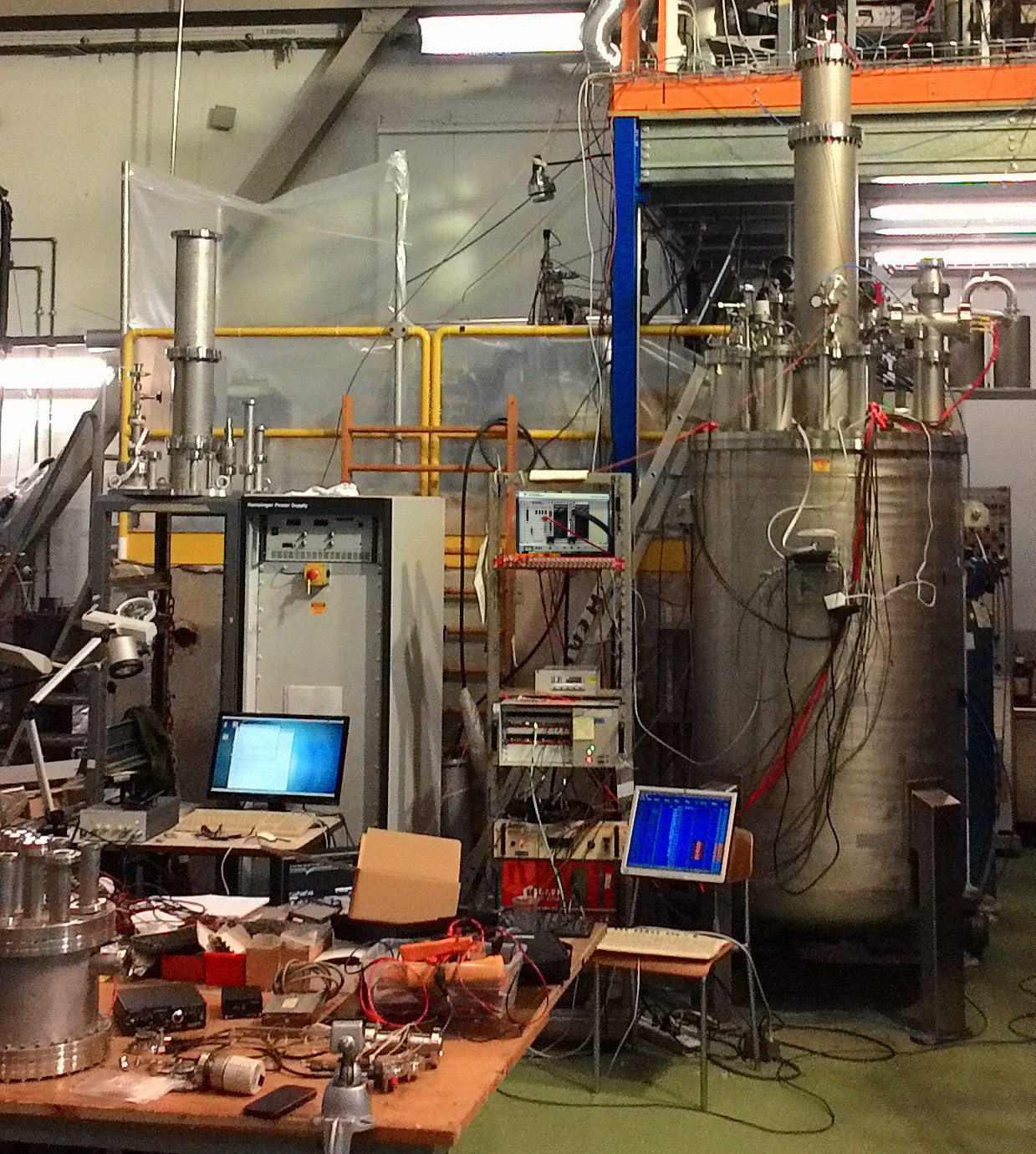}
\caption{The CERN 10-PMT facility}
\label{fig_facility}
\end{figure}

The CERN 10-PMT facility, shown in figure~\ref{fig_facility}, consists of a double-wall, vacuum-insulated 1.5~m$^3$ cryostat. The dewar is approximately 2~m high and has~112 cm external diameter, 96~cm internal diameter.  
The cryostat is internally instrumented with 10 Hamamatsu R5912-MOD PMTs with 8~in hemispherical photocathodes. Among them, 6 PMTs are coated, by means of evaporation, with about 200~$\mu$g/cm$^2$ of Tetraphenyl-Butadiene (TPB), a wavelength shifter which permits the detection of the 128~nm 
Vacuum Ultra Violet (VUV) LAr scintillation light, while 4 PMTs are left without wavelength shifter for the detection of visible photons only~\cite{Bonesini:2018ubd}. The experimental set-up is immersed in LAr. Pure commercial argon (Argon-60), certified to have purity better than 1~ppm of O$^2$ equivalent, is used in the experiment. 
During standard functioning, the system is operated in open loop, i.e. keeping it in over-pressure
with respect to the environment, and allowing LAr to evaporate.

\begin{figure}[!t]
\centering
\includegraphics[width=0.75\columnwidth]{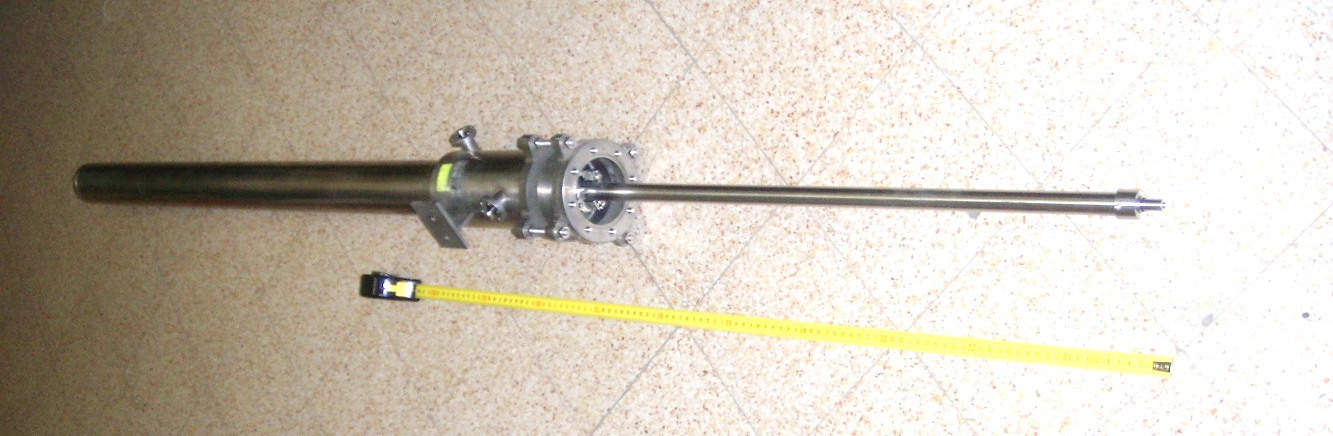}
\caption{The adopted mechanical extendible handler.}
\label{fig_translator}
\end{figure}

\begin{figure}[!t]
\centering
\includegraphics[width=0.4\columnwidth]{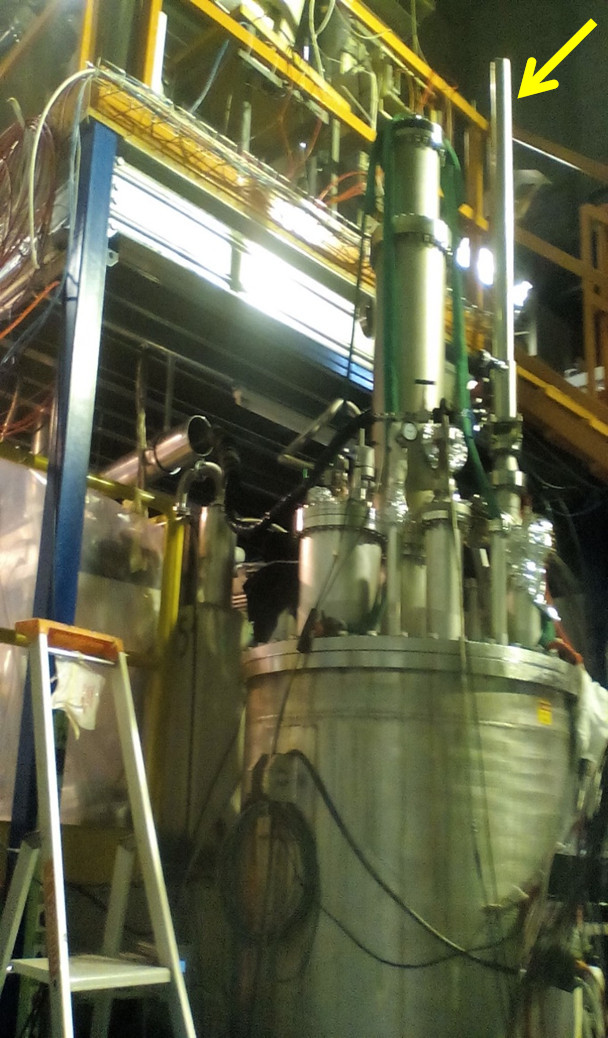}
\hspace{1cm}
\includegraphics[width=0.4\columnwidth]{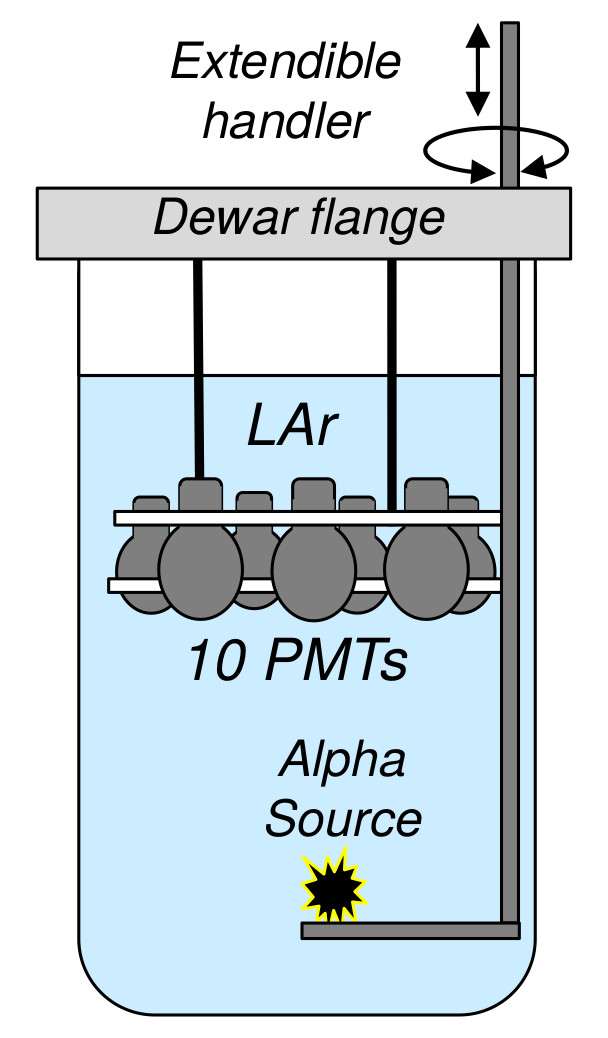}
\caption{Setup of the cryogenic facility with the mechanical extendible handler.}
\label{fig_setup}
\end{figure}

The set-up is upgraded with the installation of a mechanical extendible handler\footnote{Kenosistek S.r.l., Binasco (MI), Italy}, mounted  on the dewar top flange, which permits the internal translation and rotation of a support holding an alpha source and a SiPM-based detection system. The  mechanical handler is shown in figure~\ref{fig_translator}, while figure~\ref{fig_setup} shows a picture and a schematic of the set-up.

\begin{figure}[!t]
\centering
\includegraphics[width=0.75\columnwidth]{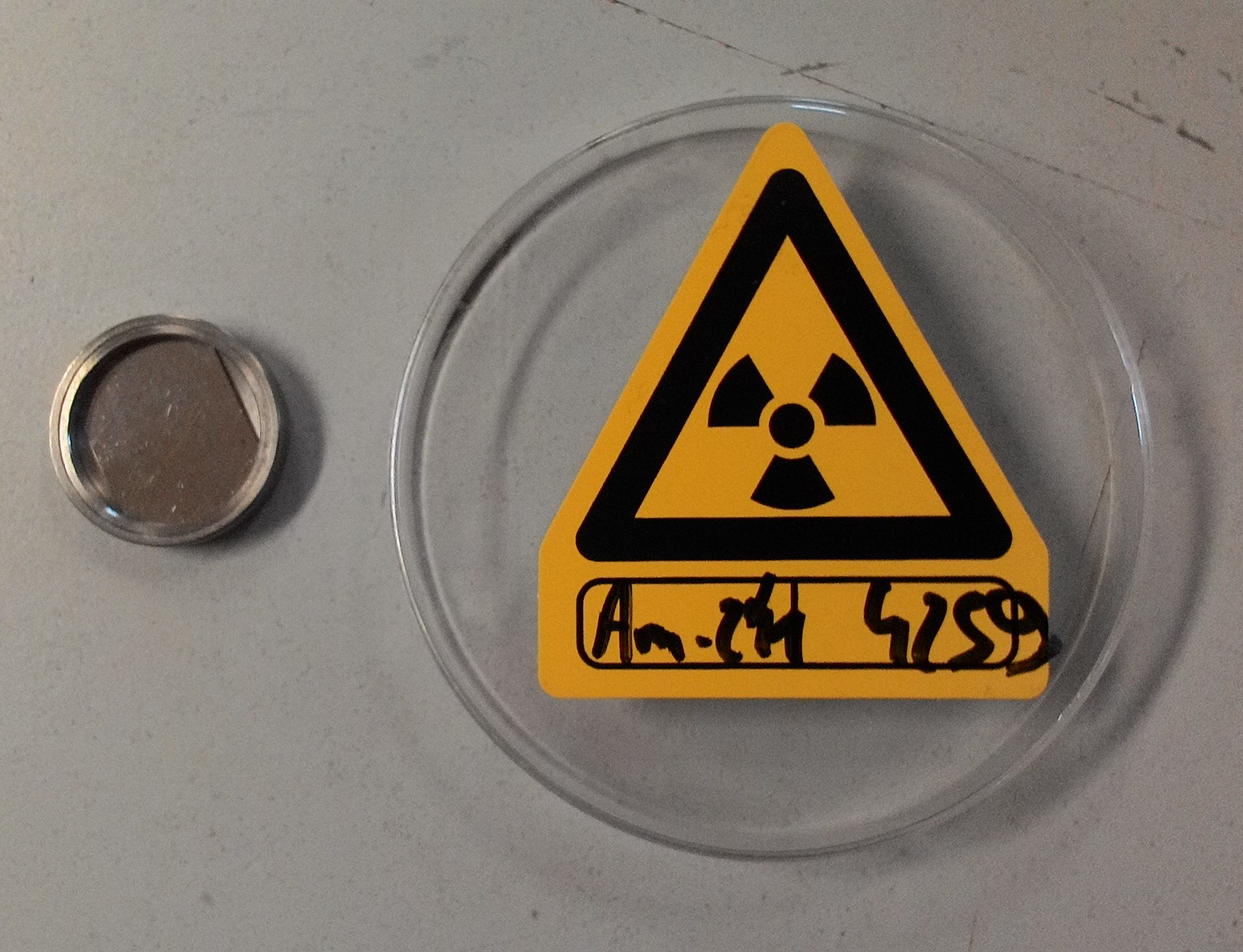}
\caption{Picture of the the used alpha source.}
\label{fig:alpha}
\end{figure}

The alpha source, shown in figure~\ref{fig:alpha}, consists of a  22~mm diameter, 4~mm thick, stainless steel disk with an active surface of about
2~cm$^2$. The activity of the source is 
39~kBq and it is able to withstand the immersion in LAr. The $^{241}$Am nuclide decays mainly via alpha emission, with energy of about 5.4~MeV.
Due to the very short stopping power of alpha particle in LAr, the light emission can be considered point-like and occurring at the 
at the same position as the source. 
This isotropic light signal propagates with negligible attenuation through the volume of the facility~\cite{Babicz:2018gqv}.
Because of their short wavelength the scintillation photons are absorbed by all detector materials without reflection, leaving time and amplitude information almost unaffected along the photon path to the light detectors. 

\begin{figure}[!t]
\centering
\includegraphics[width=0.45\columnwidth]{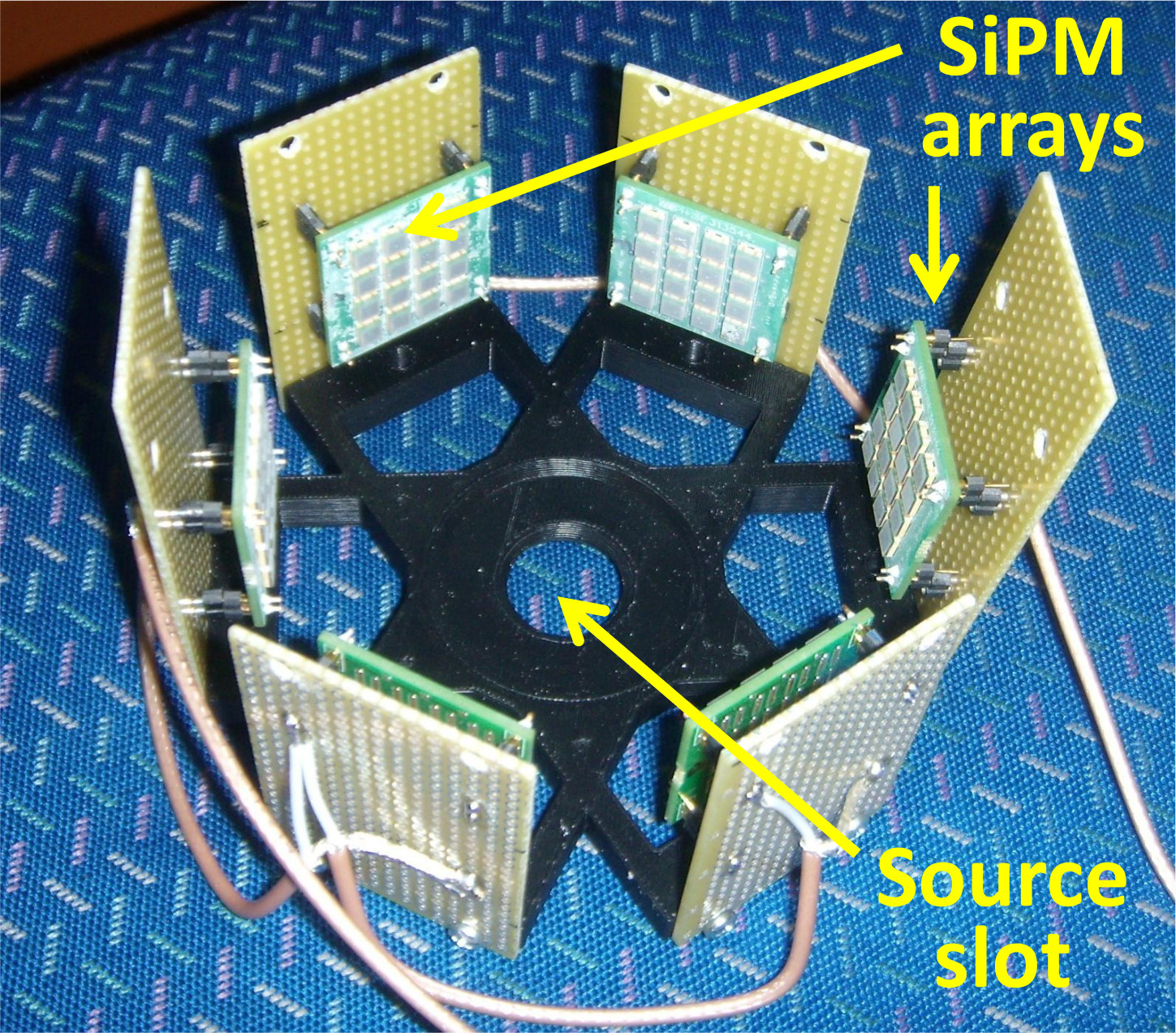}
\includegraphics[width=0.45\columnwidth]{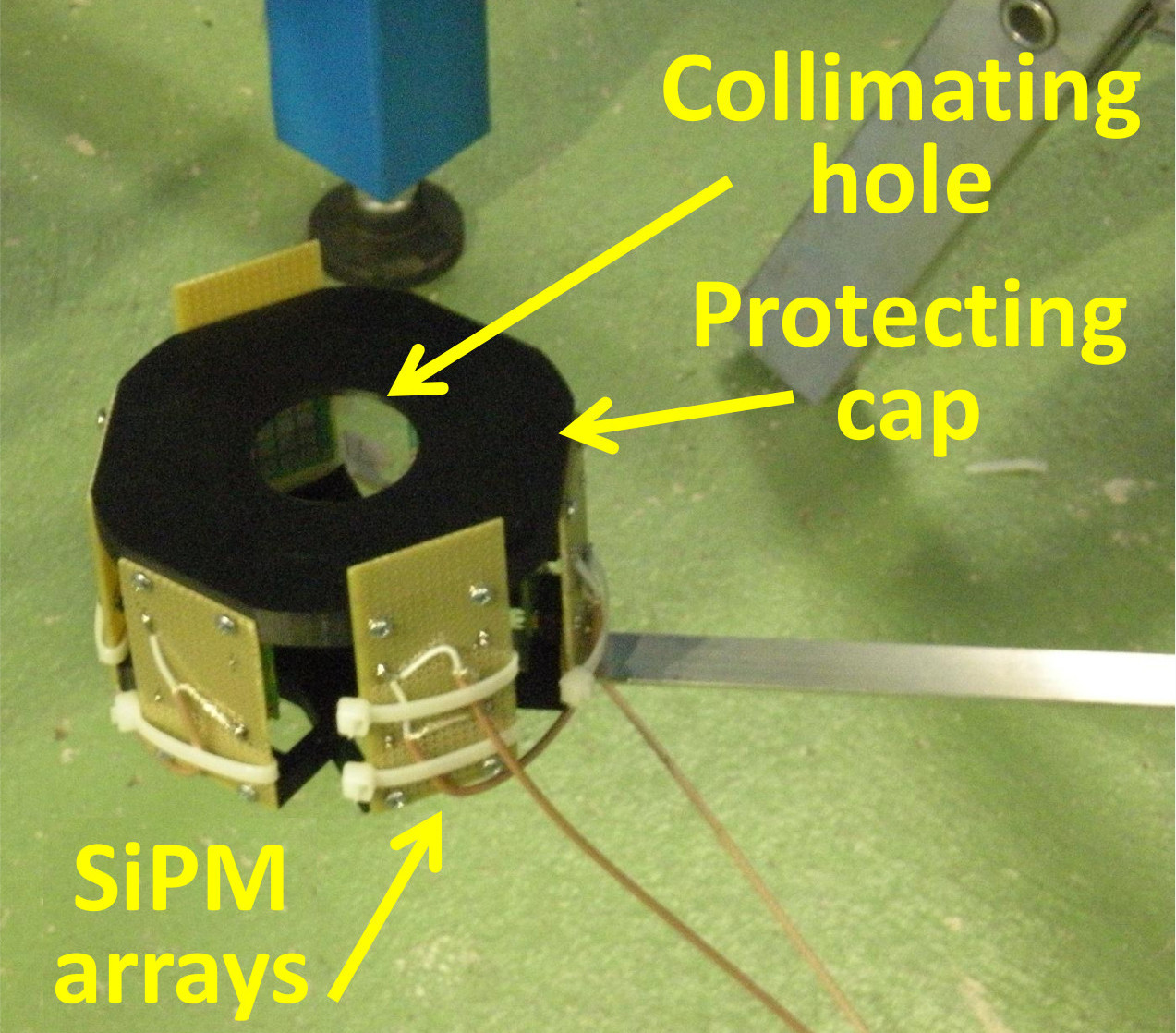}
\caption{Pictures showing the support used to handle the alpha source and the SiPM arrays.}
\label{fig:petalo}
\end{figure}

A dedicated support with hexagonal shape (110~mm width, 60~mm high), made of ABS by a 3D printer, hosts the source and supports 6 SiPM arrays (16 Hamamatsu S12572-050P for each array) used for the data acquisition trigger and for the definition of the $t_0$ of the light generation (see figure~\ref{fig:petalo}). 
In each array the devices are connected in a hybrid configuration (4 parallel arrays of 4 units in series)~\cite{CERVI2018209}.
Each array is coated with TPB to make the devices sensitive to VUV light. The 16-SiPM arrays are electrically coupled in parallel in two groups of 3 arrays, in  order to have two independent trigger lines, one of which is used as backup. Being the SiPM arrays combined with the source support, they also provide the $t_0$ signal for timing studies.
A protecting cap (110~mm diameter) with a 40~mm hole is used to shade the SiPM arrays from the direct light produced by cosmic rays in the detector active volume.

\begin{figure}[!t]
\centering
\includegraphics[width=0.75\columnwidth]{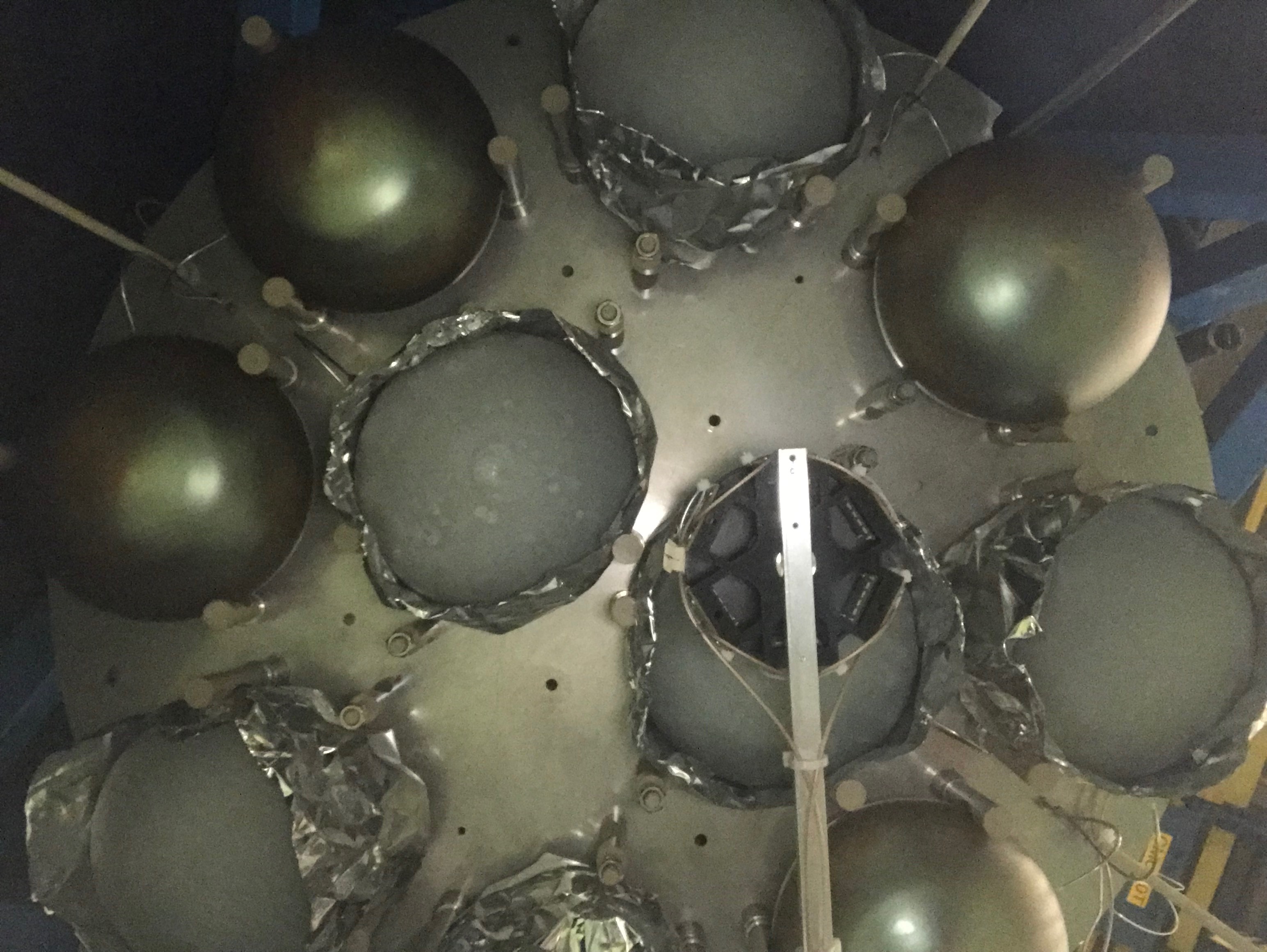}
\caption{Pictures showing the alpha source support below the PMT plane.}
\label{fig:pmt_bottom_view}
\end{figure}

The mechanical extendible handler permits to vary the source distance from the PMT surface plane in a $0 \div 90$~cm range. The rotation, is in a $0 \div 90$~degree range allowing the positioning below different PMTs as shown in figure~\ref{fig:pmt_bottom_view}.
The translation and rotation take place by means of a rack and two pinions, these last controlled by hand by means of two graduated knobs.

\section{Trigger and DAQ}

\begin{figure}[!t]
\centering
\includegraphics[width=0.45\columnwidth]{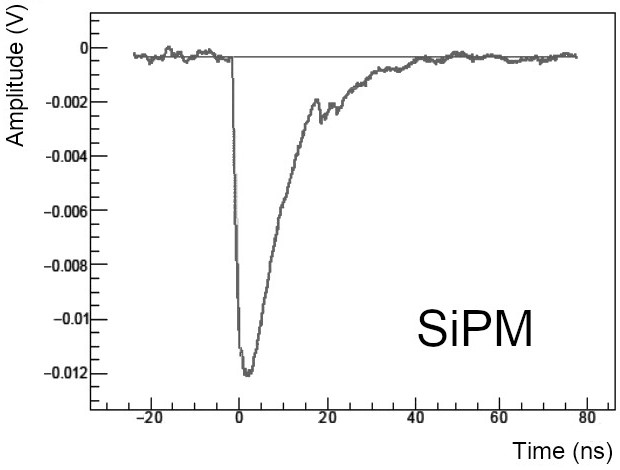}
\includegraphics[width=0.45\columnwidth]{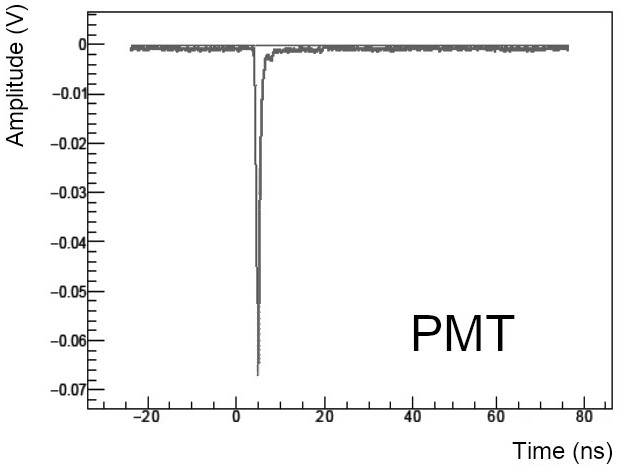}
\caption{Example of recorded signals for SiPM and PMT. Pictures show a 100~ns zoom around the trigger time.
The SiPM pulse is characterized by a trailing edge longer than that of the PMT.}
\label{fig:fit}
\end{figure}

Preliminary data are acquired 
by means of a digital oscilloscope (Tektronix MSO64, 2.5 GHz bandwidth, 12-bit 25~GSa/s) used to record for each run the signals from a SiPM array and 3 PMTs over a 1~$\mu$s window.
An example of a SiPM and a PMT signal is shown in figure~\ref{fig:fit}. 
As expected the shapes of the two signals are significantly different.
Typically SiPM pulses have longer trailing edges than PMT signals, because of the recovery time of the sensor. 
Anyway the SiPM timing properties of are not affected,
relying upon the fast (few nanoseconds) leading edge of pulses~\cite{CERVI2018209}.

\begin{figure}[!t]
\centering
\includegraphics[width=0.75\columnwidth]{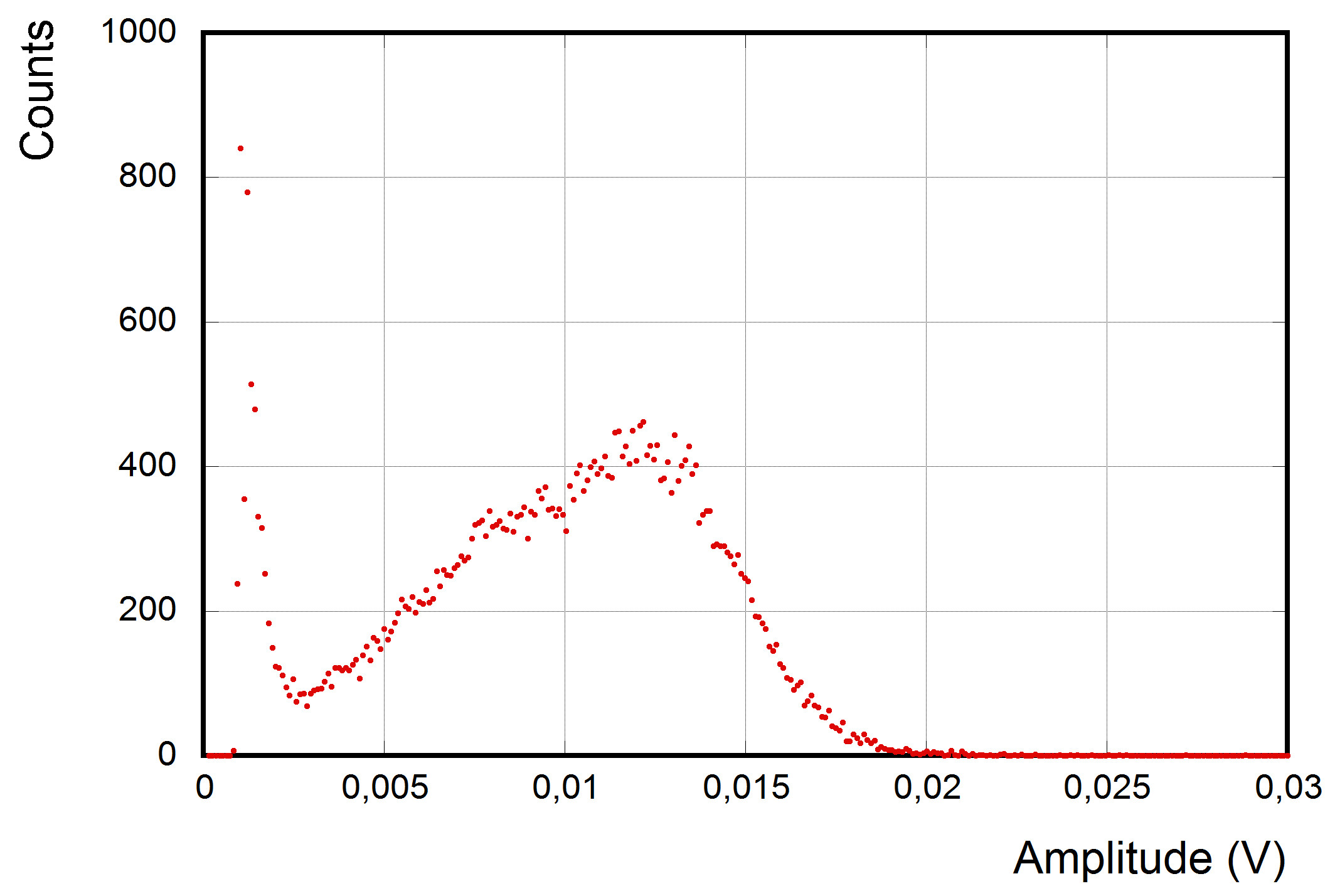}
\caption{Spectrum of alpha events recorded by one of the SiPM arrays of the setup.}
\label{fig:spectrum}
\end{figure}

The SiPM arrays voltage is set to 224~V. The corresponding amplitude distribution of the recorded pulses is shown in figure~\ref{fig:spectrum}. The PMT voltage is  set in order to have a gain of about 10$^7$. The adopted SiPM threshold level is 1~mV, while the PMT thresholds are tens of mV, corresponding to discrimination values of few photoelectrons.
The DAQ trigger is performed by the oscilloscope, requiring a coincidence in the acquisition windows of a signal above threshold from the SiPM array a signal above threshold from the PMT positioned above the source.

\section{Data analysis and system characterization}

A preliminary data set is acquired to demonstrate the effectiveness of the experimental apparatus.
In each set of acquisition runs the horizontal position of the source is set below the axis of one of the 3 recorded PMTs and this position is kept during the source vertical translation, leading to different values of delay ($t_1-t_0$) between the PMT activation and the SiPM response.

\begin{figure}[!t]
\centering
\includegraphics[width=0.75\columnwidth]{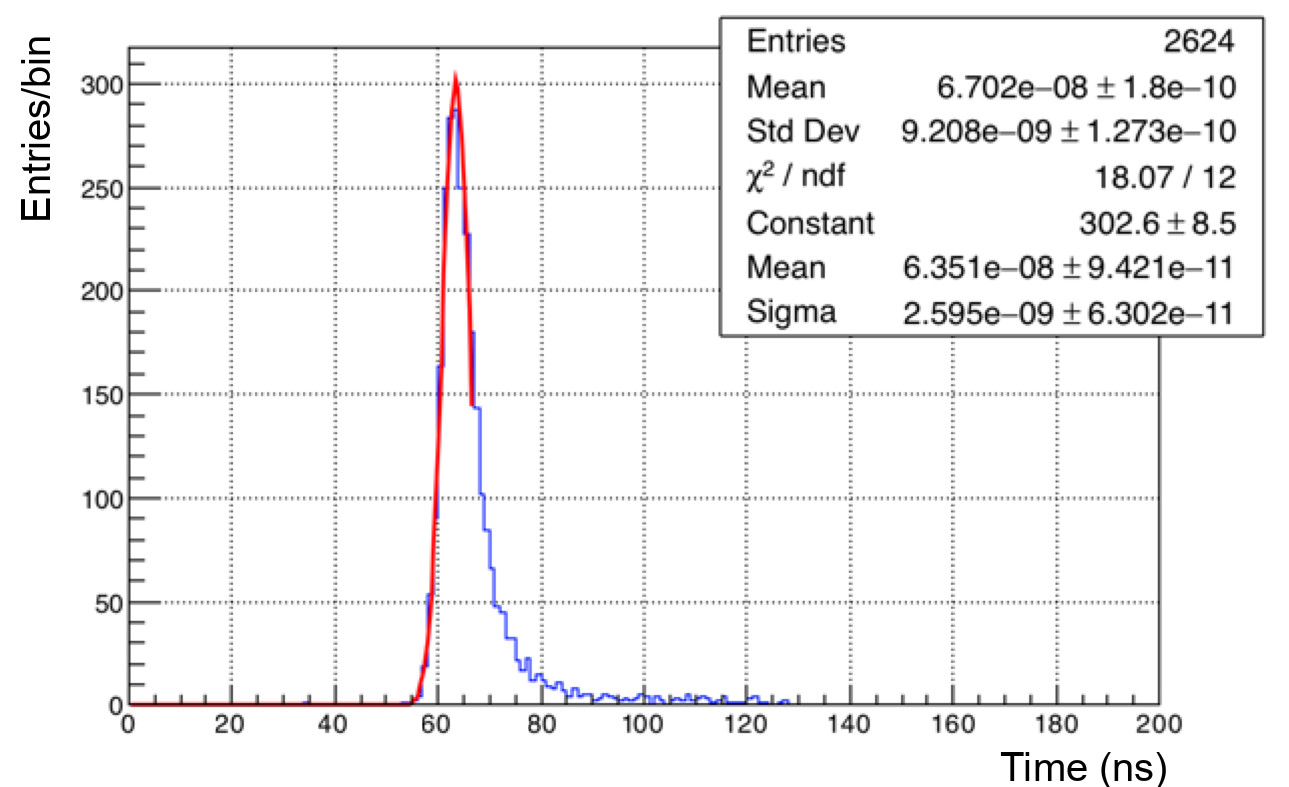}
\caption{Example of distribution of time difference between PMT and SiPM. The result of a partial Gaussian fit is shown in the figure.}
\label{fig:timedist}
\end{figure}

Two different methods are taken into account for the  determination of this delay: 1) by means of a Constant Fraction process, i.e. by measuring, event by event, the difference of the sampling points corresponding to a constant fraction (around $50$\%) 
of the leading edge of the photo-detector signals; 2) by determining for each event the arrival time of the first detected photon, i.e. the intersection point between the baseline and the line best fitting the signal leading edge.
For each source position
a distribution of delay values is obtained as shown in figure~\ref{fig:timedist}.
A slightly asymmetric shape with a tail is observed particularly for the longest distances between 
source and PMT, where the number of photons hitting the PMT is low. This is due to the 
intrinsic timing uncertainties of the PMT from small signals~\cite{burle}.
So, each distribution is only partially fitted with a Gaussian curve.

\begin{figure}[!t]
\centering
\includegraphics[width=0.75\columnwidth]{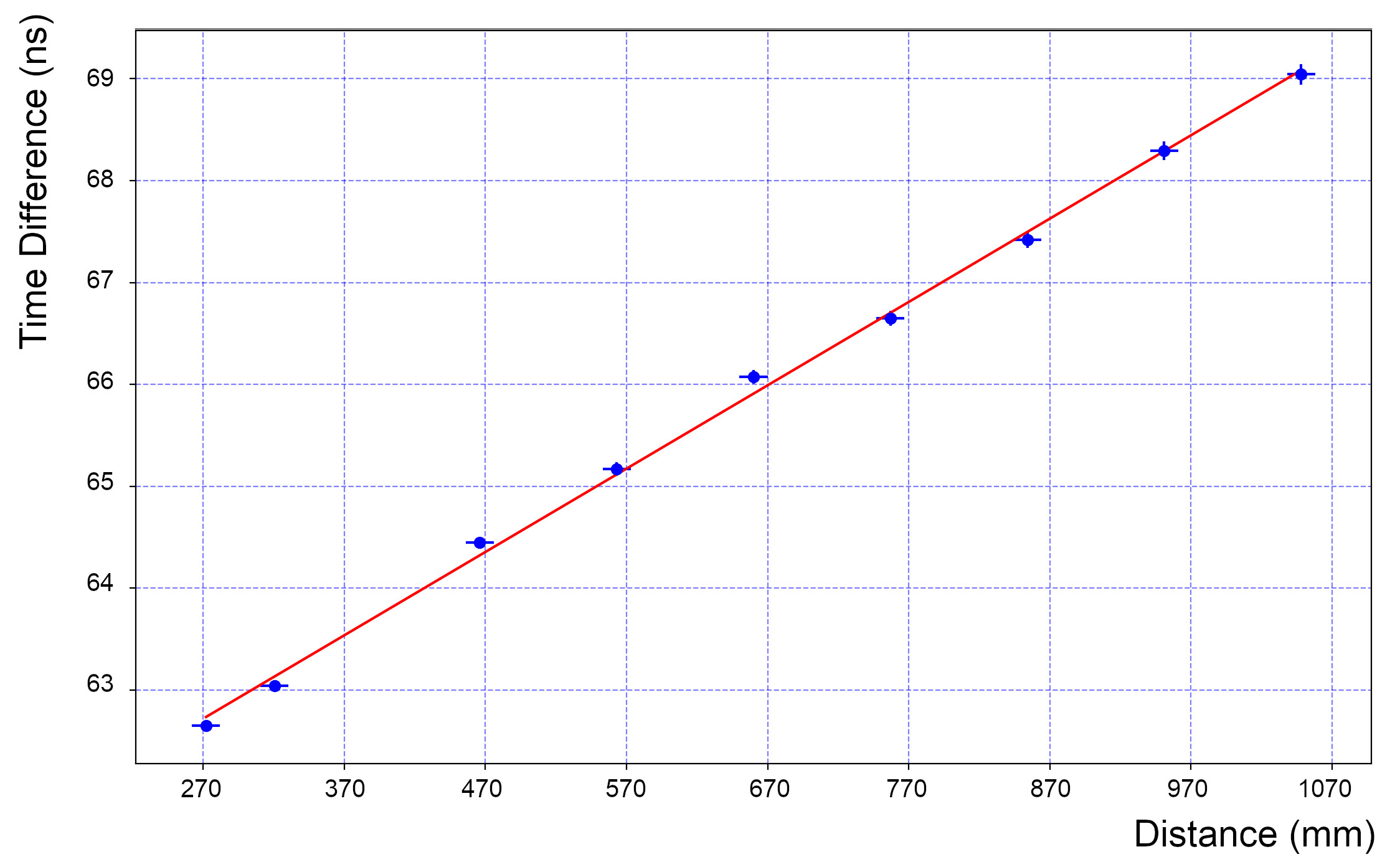}
\caption{Example of distribution of time difference between PMT and SiPM.}
\label{fig:timediff}
\end{figure}

The peak value of each distribution is plotted as a function of the distance between the SiPM and PMT to have an estimate of the delays between the response of the two photodetectors, as
shown in figure~\ref{fig:timediff}.
The slope of the straight line fitting the delays for different position gives an estimation of the velocity of VUV photons in liquid argon.
Results of measurement for the inverse of the speed of light gives values 
which strongly depend on the framework followed to determine the time differences, as well the selected PMT for the measurement.
In order to obtain valuable results,
the knowledge of all the factors that determine experimental uncertainties is required. To this purpose 
a complete Monte Carlo simulation of the facility would be useful. The precise geometry of the adopted instrumentation (shape of PMTs and SiPM, actual path of photons in active volume) and their physical characteristics, such as LAr properties and the interference of visible photons, should be taken into account.

\section{Conclusions}

A particle detection system which exploits the scintillation light produced by alpha particles 
produced by a source mounted on an extendible handler has been assembled at CERN. 
A set of 10 PMTs and 6 SiPM arrays allows the detection of the scintillation photons at different distances
between the different photodetectors.

Preliminary measurements demonstrate the capability of this system to measure
some of the properties of the LAr scintillation light. Results are still affected by systematic errors and
uncertainties which should be derived by means of a complete Monte Carlo simulation of the detector.

\section*{Acknowledgment}

The authors thank INFN and CERN Neutrino Platform for providing the necessary funding and infrastructures to perform this work.
The work of M. Babicz was supported by the National Science Center, Poland,
research project No. 2019/33/N/ST2/02874.


\bibliography{bibfile}

\end{document}